\documentclass[pra,secnumarabic,groupedaddress,showpacs]{revtex4}
\usepackage{amsmath}
\usepackage[latin1]{inputenc}

\begin{document}
\title{Representation of Joint Measurement in Quantum Mechanics\\ A Refutation of Quantum Teleportation}%
\author{Guillaume ADENIER}
\email{guillaume.adenier@ulp.u-strasbg.fr} \affiliation{Université
de Strasbourg, France.} \pacs{03.65.Ta; 03.65.Ud}
\date{October 14, 2001}
\begin{abstract}
An inconsistency is pointed out within Quantum Mechanics as soon
as successive joint measurements are involved on entangled states.
The resolution of the inconsistency leads to a refutation of the
use of entangled states as eigenvectors. Hence, the concept of
quantum teleportation, which is based on the use of such entangled
states---the Bell states---as eigenvectors, is demonstrated to be
irrelevant to Quantum Mechanics.
\end{abstract} \maketitle
\section{Introduction}
In Quantum Mechanics, the physically meaningful operators are
hermitian operators, that is, operators having only real
eigenvalues, this because a measurement performed on a single
system is always observed as a real number. However, when two
measurements are respectively performed on two distinct parts of a
whole system, the result of such a joint measurement can no longer
consist in a single real number, but in a couple of real numbers
$(a, b)$ instead, $a$ and $b$ being the results of the
measurements respectively performed on the first and second
subsystems. Yet a joint measurement is represented in Quantum
Mechanics by a tensor product observable, so that because of the
linearity of tensor product Hilbert spaces, the eigenvalue
attached to one of its eigenvectors is a single real $c$, the
computed value of the product $ab$. The point is that the
factorisation of a real $c$ into a product $ab$ is not unique, so
that two physically distinct eigenvectors can happen to have the
same eigenvalue $c$. It will be shown in this article that this
\emph{correlation degeneracy} leads to an inconsistency within
Quantum Mechanics, as soon as joint measurements are involved on
entangled states. In particular it will be shown that the concept
of quantum teleportation originates in this inconsistency.
\section{Eigenvectors and eigenvalues of tensor product
observables}\label{math}
Let $S$ be a physical system constituted
of two distinct subsystems $S_1$ and $S_2$. These subsystems are
brought arbitrarily far from each others. Let $\mathcal{H}_1$ and
$\mathcal{H}_2$ be the Hilbert spaces associated with subsystem
$S_1$ and $S_2$ respectively, so that the whole system $S$ is
represented within the tensor product space
$\mathcal{H}\equiv\mathcal{H}_1\otimes\mathcal{H}_2$.

A measurement of a variable $\mathcal{A}$, represented in Quantum
Mechanics by an observable $\hat{A}$, is to be performed on
subsystem $S_1$, while a measurement of a variable $\mathcal{B}$,
represented by an observable $\hat{B}$, is to be performed on
subsystem $S_2$. The observables $\hat{A}$ and $\hat{B}$ have both
a non degenerated spectrum as:
\begin{equation}
  \hat{A}|u_i\rangle=a_i|u_i\rangle,
\end{equation}
and
\begin{equation}
  \hat{B}|v_j\rangle=b_j|v_j\rangle.
\end{equation}
The Hilbert spaces $\mathcal{H}_1$ and $\mathcal{H}_2$ are
respectively span by the sets of eigenvectors $\{|u_i\rangle\}$
and $\{|v_j\rangle\}$, and $\mathcal{H}$ is thus spanned by the
set of tensor products of these eigenvectors
$\{|u_i\rangle\otimes|v_j\rangle\}$.

The tensor product of these observables $\hat{A}\otimes\hat{B}$ is
axiomatically defined (see Cohen-Tannoudji \emph{et al}
\cite{ctdl1}) as
\begin{equation}\label{axiometensor}
    (\hat{A}\otimes\hat{B})|\varphi\rangle\otimes|\chi\rangle
    \equiv \hat{A}|\varphi\rangle\otimes\hat{B}|\chi\rangle,
\end{equation}
where $|\varphi\rangle\otimes|\chi\rangle$ is any vector of the
tensor product Hilbert space $\mathcal{H}$. This axiomatic
definition implies straightforwardly\cite{ctdl1} that
\begin{equation}\label{tensorid}
    \hat{A}\otimes\hat{B}=
    (\hat{A}\otimes1\negmedspace\mathrm{l}_\mathrm{2})
    (1\negmedspace\mathrm{l}_\mathrm{1}\otimes\hat{B}),
\end{equation}
where $1\negmedspace\mathrm{l}_\mathrm{1}$ and
$1\negmedspace\mathrm{l}_\mathrm{2}$ are respectively the identity
operators of $\mathcal{H}_1$ and $\mathcal{H}_1$, which means that
the tensor product of observables $\hat{A}\otimes\hat{B}$ is
explicitly equivalent to a product of commuting tensor product
observables, the first one
$\hat{A}\otimes1\negmedspace\mathrm{l}_\mathrm{2}$ meaning that a
measurement represented by $\hat{A}$ is performed on subsystem
$S_1$, leaving subsystem $S_2$ unchanged, and the second one
$1\negmedspace\mathrm{l}_\mathrm{1}\otimes\hat{B}$ meaning that a
measurement represented by $\hat{B}$ is performed on subsystem
$S_2$, leaving subsystem $S_1$ unchanged.

Since it is equivalent to a product of commuting observables, this
tensor product of observables $\hat{A}\otimes\hat{B}$ is therefore
an observable too. Its eigenvectors are trivially the tensor
products of the eigenvectors of $\hat{A}$ and $\hat{B}$, that is,
\begin{eqnarray}\nonumber
  (\hat{A}\otimes\hat{B})\;|u_i\rangle\otimes|v_j\rangle
  &=&\hat{A}|u_i\rangle\otimes\hat{B}|v_j\rangle
  \\ \nonumber
  &=&a_i|u_i\rangle\otimes b_j|v_j\rangle,
\end{eqnarray}
and because of the linearity of the tensor product Hilbert space
$\mathcal{H}_1\otimes\mathcal{H}_2$, this equation can be written
as:
\begin{equation}\label{eigencomp}
  (\hat{A}\otimes\hat{B})\;|u_i\rangle\otimes|v_j\rangle=
  a_i b_j|u_i\rangle\otimes |v_j\rangle,
\end{equation}
which means that $|u_i\rangle\otimes |v_j\rangle$ is an
eigenvector of $\hat{A}\otimes\hat{B}$, and that its corresponding
eigenvalue is the product $a_ib_j$.
\subsection{Entangled eigenvectors}\label{entangeig}
However, there can exist other eigenvectors associated with the
observable $\hat{A}\otimes\hat{B}$. The point is that for any
observable, a linear combination of eigenvectors associated with
the same degenerated eigenvalue $\lambda$ is an eigenvector of
this observable as well, with the same corresponding eigenvalue
$\lambda$ (see Cohen-Tannoudji \emph{et al.}\cite{ctdl1}). Hence,
let $|u_k\rangle\otimes|v_l\rangle$ and
$|u_m\rangle\otimes|v_n\rangle$ be two distinct eigenvectors of
$\hat{A}\otimes\hat{B}$, that is,
\begin{equation}\label{distinct}
    |u_k\rangle\otimes|v_l\rangle\neq|u_m\rangle\otimes|v_n\rangle.
\end{equation}
sharing nonetheless the same degenerated eigenvalue $\lambda$ as:
\begin{equation}\label{degenerated}
  a_k b_l=a_m b_n=\lambda.
\end{equation}
This \emph{correlation degeneracy} between
$|u_k\rangle\otimes|v_l\rangle$ and
$|u_m\rangle\otimes|v_n\rangle$ implies that a linear combination
$|\psi\rangle$ of these two eigenvectors:
\begin{equation}\label{entangv}
  |\psi\rangle=\alpha_1|u_k\rangle\otimes|v_l\rangle
  +\alpha_2|u_m\rangle\otimes|v_n\rangle,
\end{equation}
is an eigenvector of $\hat{A}\otimes\hat{B}$ as well. Indeed,
\begin{eqnarray}\label{enteigenv1}
\nonumber (\hat{A}\otimes\hat{B})|\psi\rangle
    &=&\alpha_1(\hat{A}\otimes\hat{B})|u_k\rangle\otimes|v_l\rangle
      +\alpha_2(\hat{A}\otimes\hat{B})|u_m\rangle\otimes|v_n\rangle
\\ \nonumber
    &=&\alpha_1\;\lambda|u_k\rangle\otimes|v_l\rangle
      +\alpha_2\;\lambda|u_m\rangle\otimes|v_n\rangle
\end{eqnarray}
and finally:
\begin{equation}\label{enteigenv2}
  (\hat{A}\otimes\hat{B})|\psi\rangle=\lambda\;|\psi\rangle.
\end{equation}

It must be stressed that this vector $|\psi\rangle$ cannot be
factored. Suppose for instance that $|u_k\rangle=|u_m\rangle$.
Then $a_k=a_m$, so that Eq. (\ref{degenerated}) implies $b_l=b_n$,
and since the $b_j$ are non degenerated eigenvalues, it implies
$|v_l\rangle=|v_n\rangle$, and therefore
$|u_k\rangle\otimes|v_l\rangle=|u_m\rangle\otimes|v_n\rangle$
which is contradicting the above condition (\ref{distinct}). In
other words, $|\psi\rangle$ is an \emph{entangled} vector.

Therefore, Eq. (\ref{enteigenv2}) means that \emph{the entangled
vector $|\psi\rangle$ is an eigenvector of
$\hat{A}\otimes\hat{B}$}. Hence, a $\hat{A}\otimes\hat{B}$
measurement performed on a system initially in the entangled state
$|\psi\rangle$ gives with certainty the measurement result
$\lambda$, and the system remains after this measurement in the
initial state $|\psi\rangle$. In other words, the correlation
degeneracy of Eq. (\ref{degenerated}) implies that
$\hat{A}\otimes\hat{B}$ is a \emph{non demolition} measurement for
the entangled state $|\psi\rangle$. The main question that will be
addressed in this article is of the relevance of Eq.
(\ref{enteigenv2}), that is, whether it is meaningful or not to
understand an entangled vector $|\psi\rangle$ as an eigenvector of
a tensor product observable $\hat{A}\otimes\hat{B}$.
\subsection{Joint measurements on entangled vectors}\label{phys}
Although mathematically consistent with Quantum Mechanical
formalism, this non demolition prediction can actually be
questioned on a physical ground, by taking count of the local
measurement results, that is, of the fact that a
$\hat{A}\otimes\hat{B}$ measurement means that Alice performs a
$\hat{A}$ measurement on subsystem $S_1$ while Bob performs a
$\hat{B}$ measurement on subsystem $S_2$, as expressed in Eq.
(\ref{tensorid}). The point is that although the eigenvectors
$|u_k\rangle\otimes|v_l\rangle$ and
$|u_m\rangle\otimes|v_n\rangle$ of $\hat{A}\otimes\hat{B}$ do
share the same degenerated eigenvalue $\lambda$, they can
nevertheless be distinguished by the $\hat{A}\otimes\hat{B}$
measurement alone, simply by asking Alice and Bob what were
exactly the local measurements they did record each. They can
unmistakably make a distinction between
$|u_k\rangle\otimes|v_l\rangle$ and
$|u_m\rangle\otimes|v_n\rangle$, since the eigenvalues of
$\hat{A}$ and $\hat{B}$ are not degenerated (i.e., Alice can
distinguish $a_k$ from $a_m$, and Bob can distinguish $b_l$ from
$b_n$). The correlation degeneracy of the eigenvalue $\lambda$
associated with the $\hat{A}\otimes\hat{B}$ measurement is
therefore not physically meaningful since it can be resolved by
this very $\hat{A}\otimes\hat{B}$ measurement alone. Hence, the
non demolition prediction for the entangled state $|\psi\rangle$
obtained in Sect. \ref{entangeig} should be questioned as it
originates in this correlation degeneracy.

Indeed, suppose a physical system is prepared in the entangled
state $|\psi\rangle$ of Eq. (\ref{entangv}), with a normalisation
condition $|\alpha_1|^2+|\alpha_2|^2=1$, and that Alice and Bob
are to perform a $\hat{A}\otimes\hat{B}$ measurement on this
system. According to the von Neumann reduction postulate
\cite{JVN1}, the two possible joint measurement results are:
\begin{itemize}
  \item
  Alice records $a_k$ while Bob records $b_l$, in which case Alice knows
with certainty that after her $\hat{A}$ measurement the subsystem
$S_1$ is in the state $|u_k\rangle$, while Bob knows with
certainty that after his $\hat{B}$ measurement the system $S_2$ is
in the state $|v_l\rangle$, so that the state of the whole system
after the $\hat{A}\otimes\hat{B}$ measurement is
$|\psi_1'\rangle=|u_k\rangle\otimes|v_l\rangle$. The probability
  to record this joint measurement results is
\begin{equation}
  \mathcal{P}_{k,l}(\psi)=\Big|
  \big(\langle u_k|\otimes\langle v_l|\big)|\psi\rangle\Big|
  ^2=|\alpha_1|^2.
\end{equation}
  \item
  Alice records $a_m$ while Bob records $b_n$, in which case Alice knows
with certainty that after her $\hat{A}$ measurement the subsystem
$S_1$ is in the state $|u_m\rangle$, while Bob knows with
certainty that after his $\hat{B}$ measurement the system $S_2$ is
in the state $|v_n\rangle$, so that the state of the whole system
after the $\hat{A}\otimes\hat{B}$ measurement is
$|\psi_2'\rangle=|u_m\rangle\otimes|v_n\rangle$. The probability
  to record this joint measurement results is
\begin{equation}
  \mathcal{P}_{m,n}(\psi)=\Big|
  \big(\langle u_m|\otimes\langle v_n|\big)|\psi\rangle\Big|
  ^2=|\alpha_2|^2.
\end{equation}
\end{itemize}
Hence, a $\hat{A}\otimes\hat{B}$ measurement performed on an
entangled system $|\psi\rangle$ transforms it either into a system
represented by the state
$|\psi_1'\rangle=|u_k\rangle\otimes|v_l\rangle$ or into a system
represented by the state
$|\psi_2'\rangle=|u_m\rangle\otimes|v_n\rangle$, depending on the
measurement results recorded by Alice and Bob, which means that
the $\hat{A}\otimes\hat{B}$ measurement is a \emph{demolition}
measurement for the entangled state $|\psi\rangle$, contrary to
what was suggested by Eq. (\ref{enteigenv2}). Therefore, the use
of correlation degeneracy, which is responsible for the incorrect
understanding of the entangled state $|\psi\rangle$ as an
eigenvector of $\hat{A}\otimes\hat{B}$, is physically meaningless
and must be discarded.
\subsection{Joint Eigenvalues}\label{jointeigen}
The correlation degeneracy originates in the linearity of tensor
product Hilbert spaces, that is, in
\begin{equation}\label{linearity}
    \alpha|\varphi\rangle\otimes\beta|\chi\rangle=
    (\alpha\beta)|\varphi\rangle\otimes|\chi\rangle,
\end{equation}
where $\alpha$ and $\beta$ are any complex numbers. As seen in
Sect. \ref{phys}, when these numbers are eigenvalues, this
computation of two localized real eigenvalues into a non-localized
single eigenvalue $\lambda$ makes no physical sense since a part
of the information is arbitrarily lost in the process, although
this loss has no physical counterpart. Thus, in order to get only
physically meaningful predictions, such as those of Sect.
\ref{phys}, the proposal of this article is to slightly change
Quantum Mechanical formalism for joint measurements, so that a
joint measurement can no longer be associated with only one single
computed eigenvalue $\lambda$, but with a couple of eigenvalues
$(\alpha,\beta)$ instead, that is, a \emph{joint eigenvalue}
axiomatically defined as:
\begin{equation}
  (\alpha,\beta)|\varphi\rangle\otimes|\chi\rangle\equiv
  \alpha|\varphi\rangle\otimes\beta|\chi\rangle,
\end{equation}
so that the eigenvectors of $\hat{A}\otimes\hat{B}$ are such that
\begin{equation}
    (\hat{A}\otimes\hat{B})|u_i\rangle\otimes|v_j\rangle=
    (a_i,b_j)|u_i\rangle\otimes|v_j\rangle,
\end{equation}
which reads as follow: $(a_i,b_j)$ is the joint eigenvalue
associated with the eigenvector $|u_i\rangle\otimes|v_j\rangle$.
The advantage of this notation is that the physically distinct
eigenvectors $|u_k\rangle\otimes|v_l\rangle$ and
$|u_m\rangle\otimes|v_n\rangle$ are no longer associated with the
same degenerated eigenvalue $\lambda$, but with distinct joint
eigenvalue instead:
\begin{subequations}
\begin{eqnarray}
    (\hat{A}\otimes\hat{B})|u_k\rangle\otimes|v_l\rangle&=&
    (a_k,b_l)|u_k\rangle\otimes|v_l\rangle
    \\
    (\hat{A}\otimes\hat{B})|u_m\rangle\otimes|v_n\rangle&=&
    (a_m,b_n)|u_m\rangle\otimes|v_n\rangle,
\end{eqnarray}
\end{subequations}
with $(a_k,b_l)\neq(a_m,b_n)$, meaning that these joint
eigenvalues are non degenerated, as $\{a_i\}$ and $\{b_j\}$ are
non degenerated spectrum.

With this explicit notation, the physically meaningless loss of
information due to the use of correlation degeneracy is avoided,
and the entangled state $|\psi\rangle$ no longer appears
deceitfully as an eigenvector of $\hat{A}\otimes\hat{B}$ since the
eigenvectors it is made of are associated with distinct joint
eigenvalues. This allows conveniently to recover directly the
correct physical predictions given in Sect. \ref{phys}, without
allowing the wrong predictions of Sect. \ref{entangeig}.

Unfortunately, these wrong predictions based on correlation
degeneracy seem at first sight so mathematically correct that
there exist many examples in the recent literature of such an
implicit misuse of Quantum Mechanical formalism, among which are
the famous Bell states and Quantum Teleportation.
\section{Application to the Bell states}\label{Appl}
The Bell states\cite{Braunstein1} are entangled states defined
within tensor products of spin Hilbert spaces. Spin measurements
can be represented within spin Hilbert space by the Pauli matrices
$\sigma_x$, $\sigma_y$, and $\sigma_z$ (to simplify notation,
${\hbar}/{2}$ is set to $1$ throughout), with the following
eigenvectors and eigenvalues:
\begin{eqnarray}\label{paulimatr}
    \sigma_z & |\pm\rangle &=\pm |\pm\rangle \\
    \sigma_x & |\pm\rangle_x &=\pm |\pm\rangle_x \\
    \sigma_y & |\pm\rangle_y &=\pm |\pm\rangle_y.
\end{eqnarray}
and the eigenvectors of $\sigma_x$ and $\sigma_y$ can be written
within the $z$-basis $\{|+\rangle,|-\rangle\}$ as
\begin{eqnarray}\label{spinx}
    |\pm\rangle_x&=&\frac{1}{\sqrt{2}}\Big[|+\rangle\pm|-\rangle\Big]
    \\
    |\pm\rangle_y&=&\frac{1}{\sqrt{2}}\Big[|+\rangle\pm i
    |-\rangle\Big].
\end{eqnarray}
The tensor product Hilbert space
$\mathcal{H}=\mathcal{H}_1\otimes\mathcal{H}_2$ where the Bell
states are to be written can thus be spanned by the product basis
formed by the four eigenvectors $\{ |++\rangle, |+-\rangle,
|-+\rangle, |--\rangle \}$ associated with the tensor product
observable $\sigma_z\otimes\sigma_z$.
\subsection{Bell states with correlation
degeneracy}\label{Bellcode}
With the use of correlation
degeneracy, the eigenvector of $\sigma_z\otimes\sigma_z$ have two
possible degenerated eigenvalues $-1$ or $+1$ as:
\begin{subequations}
    \label{Sigmaeigen}
\begin{eqnarray}
\label{sigmaea} \sigma_z\otimes\sigma_z|++\rangle&=&+|++\rangle\\
\label{sigmaeb} \sigma_z\otimes\sigma_z|+-\rangle&=&-|+-\rangle\\
\label{sigmaec} \sigma_z\otimes\sigma_z|-+\rangle&=&-|-+\rangle\\
\label{sigmaed} \sigma_z\otimes\sigma_z|--\rangle&=&+|--\rangle,
\end{eqnarray}
\end{subequations}
that is, on one hand $|+-\rangle$ and $|-+\rangle$ are
eigenvectors associated with the degenerated eigenvalue $-1$, and
on the other hand $|++\rangle$ and $|--\rangle$ are eigenvectors
associated with the degenerated eigenvalue $+1$ (see Eq.
\ref{degenerated}).

Hence, akin to what was done in Sect. \ref{entangeig}, one can
build other eigenvectors of $\sigma_z\otimes\sigma_z$, by
combining eigenvectors associated with the same degenerated
eigenvalue (see Eq. \ref{entangv}), that is, two normalized states
$|\Psi^+\rangle$ and $|\Psi^-\rangle$ made up with eigenvectors
(\ref{sigmaeb}) and (\ref{sigmaec}) associated with the
degenerated eigenvalue $-1$:
\begin{subequations}
\begin{eqnarray}
    |\Psi^+\rangle&\equiv&\frac{1}{\sqrt{2}}\Big[|+-\rangle+|-+\rangle\Big],
\\
    |\Psi^-\rangle&\equiv&\frac{1}{\sqrt{2}}\Big[|+-\rangle-|-+\rangle\Big],
    \label{singlet}
\end{eqnarray}
\end{subequations}
and two others normalized states $|\Phi^+\rangle$ and
$|\Phi^-\rangle$ made up with eigenvectors (\ref{sigmaea}) and
(\ref{sigmaed}) associated with the degenerated eigenvalue $+1$ :
\begin{subequations}
\begin{eqnarray}
  |\Phi^+\rangle&\equiv&\frac{1}{\sqrt{2}}\Big[|++\rangle+|--\rangle\Big],
\\
  |\Phi^-\rangle&\equiv&\frac{1}{\sqrt{2}}\Big[|++\rangle-|--\rangle\Big].
\end{eqnarray}
\end{subequations}
These are the so called \emph{Bell states}\cite{Braunstein1}, and
one can verify easily that these vectors are orthonormal
\cite{Bennet1}, so that they can apparently be used as basis
vectors to span the tensor product Hilbert space $\mathcal{H}$.

Following the argument of Sect. \ref{entangeig}, one can make use
of the fact that a Bell state is built up exclusively with
eigenvectors of $\sigma_z\otimes\sigma_z$ associated with one
degenerated eigenvalue, so that it is an eigenvectors of
$\sigma_z\otimes\sigma_z$ as well, with this same eigenvalue (see
Eq. \ref{enteigenv2}):
\begin{subequations}\label{Bell1}
\begin{eqnarray}\label{singleteigz}
  \sigma_z\otimes\,\sigma_z|\Psi^\pm\rangle
  &=&-|\Psi^\pm\rangle
\\
  \sigma_z\otimes\,\sigma_z|\Phi^\pm\rangle
  &=&+|\Phi^\pm\rangle,
\end{eqnarray}
\end{subequations}
and with the help of Eq. (\ref{spinx}), the Bell states can be
shown to be eigenvectors of the observable
$\sigma_x\otimes\sigma_x$ too:
\begin{subequations}\label{Bell2}
\begin{eqnarray}\label{singleteigx}
  \sigma_x\otimes\,\sigma_x|\Psi^\pm\rangle
  &=&\pm|\Psi^\pm\rangle,
\\
   \sigma_x\otimes\,\sigma_x|\Phi^\pm\rangle
  &=&\pm|\Phi^\pm\rangle,
\end{eqnarray}
\end{subequations}
(See Cabello \cite{Cabello1} for instance), which means that
$\sigma_z\otimes\sigma_z$ and $\sigma_x\otimes\sigma_x$ are
\emph{non demolition} operators for the Bell states.

An important consequence is that for any state $|\psi\rangle$
written within the Bell state basis, the following equation holds:
\begin{equation}\label{commutation}
  (\sigma_z\otimes\,\sigma_z)(\sigma_x\otimes\,\sigma_x)|\psi\rangle
  =(\sigma_x\otimes\,\sigma_x)(\sigma_z\otimes\,\sigma_z)|\psi\rangle
\end{equation}
as one can easily verify with Eqs. (\ref{Bell1}) and
(\ref{Bell2}), which means that $\sigma_z\otimes\sigma_z$ and
$\sigma_x\otimes\sigma_x$ are \emph{commuting observables}:
\begin{equation}\label{comm}
    \big[\sigma_z\otimes\sigma_z,\sigma_x\otimes\sigma_x\Big]
    =0,
\end{equation}

However, these widely accepted results \cite{Braunstein1,
Cabello1, Braunstein2, Larsen1, Cabello2, Aravind1, Lomonaco1} are
based on the use of correlation degeneracy. Since this use was
shown in Sect. \ref{phys} to lead to wrong physical predictions,
and thus discarded, a closer analysis with the help of joint
eigenvalues is required.
\subsection{Bell states with joint eigenvalues}\label{Belljoei}
The joint eigenvalues associated with the eigenvectors of
$\sigma_z\otimes\sigma_z$ are as follow:
\begin{subequations}
    \label{Sigmajeigen}
\begin{eqnarray}
    \sigma_z\otimes\sigma_z |++\rangle&=&(+1,+1)|++\rangle\\
    \sigma_z\otimes\sigma_z |+-\rangle&=&(+1,-1)|+-\rangle\\
    \sigma_z\otimes\sigma_z |-+\rangle&=&(-1,+1)|-+\rangle\\
    \sigma_z\otimes\sigma_z |--\rangle&=&(-1,-1)|--\rangle,
\end{eqnarray}
\end{subequations}
These distinct joint eigenvalues make explicit the fact that Alice
and Bob can unmistakably discriminate these states by means of the
$\sigma_z\otimes\sigma_z$ measurement alone (see Sect. \ref{phys})
since they can both make a distinction between a $+1$ result and a
$-1$ result. The eigenvectors of $\sigma_z\otimes\sigma_z$ are
therefore no longer associated with degenerated eigenvalues, but
with \emph{non degenerated joint eigenvalues} instead, so that
they can no longer be combined to form other eigenvectors. Hence,
the Bell states no longer appear deceitfully as eigenvectors of
$\sigma_z\otimes\sigma_z$, $\sigma_x\otimes\sigma_x$, or
$\sigma_y\otimes\sigma_y$. In other words, the Eqs. (\ref{Bell1}),
(\ref{Bell2}), and (\ref{commutation}) which are all based on the
use of correlation degeneracy are therefore incorrect.

To pinpoint this more explicitly, suppose Alice and Bob are to
perform a $\sigma_z\otimes\sigma_z$ measurement on a system
initially in the singlet state $|\Psi^-\rangle$ of Eq.
(\ref{singlet}). In this state, $|+-\rangle$ corresponds to the
joint eigenvalue $(+1,-1)$, whereas $|-+\rangle$ corresponds to
the joint eigenvalue $(-1,+1)$. Hence, according to the projection
postulate, the measurement results recorded by Alice and Bob are
either given by:
\begin{itemize}
  \item
the joint eigenvalue $(+1,-1)$, meaning that Alice records $+1$
while Bob records $-1$, in which case Alice knows with certainty
that after her $\sigma_z$ measurement the subsystem $S_1$ is in
the state $|+\rangle$, while Bob knows with certainty that after
his $\sigma_z$ measurement the system $S_2$ is in the state
$|-\rangle$, so that the state of the whole system after the
$\sigma_z\otimes\sigma_z$ measurement is
$|\psi_1'\rangle=|+-\rangle$. The probability to record this joint
measurement result is
\begin{equation}
    \mathcal{P}_{+-}=\Big|\langle+-|\Psi^-\rangle\Big|^2=\frac{1}{2},
\end{equation}
  \item
the joint eigenvalue $(-1,+1)$, meaning that Alice records $-1$
while Bob records $+1$, in which case Alice knows with certainty
that after her $\sigma_z$ measurement the subsystem $S_1$ is in
the state $|-\rangle$, while Bob knows with certainty that after
his $\sigma_z$ measurement the system $S_2$ is in the state
$|+\rangle$, so that the state of the whole system after the
$\sigma_z\otimes\sigma_z$ measurement is
$|\psi_2'\rangle=|-+\rangle$. The probability to record this joint
measurement result is
\begin{equation}
    \mathcal{P}_{-+}=\Big|\langle-+|\Psi^-\rangle\Big|^2=\frac{1}{2}.
\end{equation}
\end{itemize}
Once this $\sigma_z\otimes\sigma_z$ measurement has been
performed, \emph{the system is no longer in the singlet state}, it
is either in the state $|\psi_1'\rangle=|+-\rangle$ or in the
state $|\psi_2'\rangle=|-+\rangle$, depending on the measurement
results recorded by Alice and Bob. The $\sigma_z\otimes\sigma_z$
measurement is therefore a \emph{demolition} measurement for the
singlet state $|\Psi^-\rangle$, unlike what was found in Sect.
\ref{Bellcode}.

This has important consequences. For instance, let's suppose the
latter case $(-1,+1)$ has been realized : Alice did record a $-1$
result, while Bob recorded $+1$. The state of the system after
this measurement is therefore $|\psi'_2\rangle=|-+\rangle$. Now,
Alice and Bob are to perform the second measurement
$\sigma_x\otimes\sigma_x$ on this system. The state of the system
must therefore be projected onto the $x$-basis with the help of
Eq. (\ref{spinx}), that is,
\begin{equation}\label{jointpr}
  |\psi_2'\rangle=\frac{1}{2}\Big[|++\rangle_x-|+-\rangle_x+|-+\rangle_x-|--\rangle_x\Big],
\end{equation}
which means that there are four possible couples of outcomes:
$(+1,+1)$, $(+1,-1)$, $(-1,+1)$, or $(-1,-1)$, each joint
eigenvalue having the same probability $\frac{1}{4}$ to occur.
Note that the correlation between Alice and Bob measurement
results is no longer necessarily equal to $-1$, unlike what would
have been predicted using Eqs. (\ref{singleteigz}) and
(\ref{singleteigx}). After this second measurement the system is
in the $x$-basis vectors $|++\rangle_x$, $|+-\rangle_x$,
$|-+\rangle_x$ or $|--\rangle_x$ corresponding to the observed
joint eigenvalue. Had the measurements been performed in the
reverse order, that is, the $\sigma_x\otimes\sigma_x$ measurement
followed by the $\sigma_z\otimes\sigma_z$ measurement, the final
state of the system would have been one of the $z$-basis vectors
$|++\rangle$, $|+-\rangle$, $|-+\rangle$, or $|--\rangle$. Since
any $x$-basis vector is distinct from any $z$-basis vector, then
$(\sigma_x\otimes\sigma_x)(\sigma_z\otimes\sigma_z)$ and
$(\sigma_z\otimes\sigma_z)(\sigma_x\otimes\sigma_x)$ are not at
all equivalent measurements, which means that
$\sigma_z\otimes\sigma_z$ and $\sigma_x\otimes\sigma_x$ \emph{are
not commuting observables}:
\begin{equation}\label{nocomm}
    \big[\sigma_z\otimes\sigma_z,\sigma_x\otimes\sigma_x\Big]
    \neq 0,
\end{equation}
contrary to what was found in Sect. \ref{Bellcode} and most unlike
what can be found in the literature. Note that since the incorrect
commutation prediction of Eq. (\ref{comm}) was due to the implicit
use of correlation degeneracy, the use of joint eigenvalues would
have directly prevented from such a mistake. Incidentally, this
non commutation result based on quantum postulates is in itself
sufficient to refute a class of Bell's theorem without
inequalities \cite{Cabello1, Aravind1,Mermin1} based on this
physically incorrect commutation of $\sigma_z\otimes\sigma_z$ and
$\sigma_x\otimes\sigma_x$, or similarly of
$\sigma_z\otimes\sigma_x$ and $\sigma_x\otimes\sigma_z$.
\section{Application to Quantum Teleportation}\label{QTeleport}
\subsection{Quantum Teleportation Scheme}\label{QTscheme}
One of the most important use of the Bell states is made by
Quantum Teleportation, as originally conceived by Bennet \emph{et
al}\cite{Bennet1}. Basically, the principle of Quantum
Teleportation is as follow:

\begin{description}
  \item [Step 1 -- Preparation:]

Alice and Bob must share beforehand a system constituted of two
particles, labeled 2 and 3, prepared in a singlet state and thus
written within the product space
$\mathcal{H}_2\otimes\mathcal{H}_3$ as:
\begin{equation}\label{Ancil}
  |\Psi^-_{23}\rangle=\frac{1}{\sqrt{2}}\Big[
  |+\rangle\otimes|-\rangle-|-\rangle\otimes|+\rangle
  \Big].
\end{equation}
Alice takes particle 2 with her, and Bob goes wherever he wants
with particle 3 as long as particles 2 and 3 remain entangled.
Then, Alice is given an unknown state carried by a particle,
labeled 1, written within a spin Hilbert space $\mathcal{H}_1$ as:
\begin{equation}\label{unknownst}
|\phi_1\rangle=a|+\rangle+b|-\rangle,
\end{equation}
so that the state of the whole system owned by Alice and Bob is:
\begin{equation}\label{init}
  |\psi_{123}\rangle=|\phi_1\rangle\otimes|\Psi^-_{23}\rangle.
\end{equation}
Alice's purpose is to have the state carried by particle 1
``teleported" to particle 3.

  \item [Step 2 -- Projection within Bell states basis:]

The state of the whole system, initially written within the spin
states basis is projected onto the Bell states basis using the
Bell identity:
\begin{equation}\label{identBell}
  \hat{1\negmedspace\mathrm{l}}_\mathrm{Bell}=
  |\Psi^-\rangle\langle\Psi^-|+|\Psi^+\rangle\langle\Psi^+|
  +|\Phi^-\rangle\langle\Phi^-|+|\Phi^+\rangle\langle\Phi^+|,
\end{equation}
that is,
\begin{eqnarray}\label{TelEq1}\nonumber
    (\hat{1\negmedspace\mathrm{l}}_\mathrm{Bell}
    \otimes\hat{1\negmedspace\mathrm{l}}_3)
    |\psi_{123}\rangle=
    \frac{1}{2}\Big[
    |\Psi^-_{12}\rangle\otimes\Big(-a|+\rangle-b|-\rangle\Big)
   +|\Psi^+_{12}\rangle\otimes\Big(-a|+\rangle+b|-\rangle\Big)\\
   +|\Phi^-_{12}\rangle\otimes\Big(a|+\rangle+b|-\rangle\Big)
   +|\Phi^+_{12}\rangle\otimes\Big(a|+\rangle-b|-\rangle\Big)
   \Big].
\end{eqnarray}
Note that at this point no physical modification of the system has
occurred yet.

    \item [Step 3 -- Reduction of the pure state:]

The previous equation suggests to find a \emph{measurement}
capable of collapsing the \emph{pure} state of the system
$|\psi_{123}\rangle$ into either one of the following states:
\begin{subequations}
    \label{TelEq2}
\begin{eqnarray}
    |\psi'_a\rangle&=&|\Psi^-_{12}\rangle\otimes\Big(-a|+\rangle-b|-\rangle\Big),\\
    |\psi'_b\rangle&=&|\Psi^+_{12}\rangle\otimes\Big(-a|+\rangle+b|-\rangle\Big),\\
    |\psi'_c\rangle&=&|\Phi^-_{12}\rangle\otimes\Big(a|+\rangle+b|-\rangle\Big),\\
    |\psi'_d\rangle&=&|\Phi^+_{12}\rangle\otimes\Big(a|+\rangle-b|-\rangle\Big).
\end{eqnarray}
\end{subequations}
Once this reduction has occurred, the exact state of the system is
still unknown to Alice and Bob so that the density matrix
representing the system is the \emph{mixture}:
\begin{equation}\label{density1}
  \rho'_{123}=\frac{1}{4}\Big[
   |\psi'_a\rangle\langle\psi'_a|
  +|\psi'_b\rangle\langle\psi'_b|
  +|\psi'_c\rangle\langle\psi'_c|
  +|\psi'_d\rangle\langle\psi'_d|
  \Big].
\end{equation}
This reduction is the crucial step of the whole process. This
point will be reviewed with more details in the next section.

  \item [Step 4 -- Disentanglement:]
  Alice must determine which of the states of Eqs. (\ref{TelEq2}) is actually the state
  of the system. For this purpose, Alice must disentangle particles 1 and 2 in order
  to be able to ascribe a definite state to each particle by means of local operations.
  The disentanglement is performed by a a conditional spin flip \cite{Vaidman1},
  a unitary transformation $\hat{U}_C$ defined as:
\begin{equation}\label{cnot}
  \hat{U}_C=
   |+-\rangle\langle++|
  \;+\;|++\rangle\langle+-|
  \;+\;|-+\rangle\langle-+|
  \;+\;|--\rangle\langle--|,
\end{equation}
which is a quantum-quantum interaction between $S_1$ and $S_2$.
The state of the system after this transformation is:
\begin{eqnarray}\label{transdens}
    \rho''_{123}&=&(\hat{U}_C\otimes\hat{1\negmedspace\mathrm{l}}_3)
    \;\rho'_{123}\;
    (\hat{U}^\dag_C\otimes\hat{1\negmedspace\mathrm{l}}_3)\\
    \label{transdens2}
                &=&\frac{1}{4}\sum_\alpha
    (\hat{U}_C\otimes\hat{1\negmedspace\mathrm{l}}_3)
    |\psi'_\alpha\rangle\langle\psi'_\alpha|
    (\hat{U}_C\otimes\hat{1\negmedspace\mathrm{l}}_3)
\end{eqnarray}
which means explicitly that the system is in a mixture of the
following pure states:
\begin{subequations}\label{TelEq3b}
\begin{eqnarray}
    |\psi''_a\rangle&=&\frac{1}{\sqrt{2}}
    \Big(|+\rangle-|-\rangle\Big)\otimes|+\rangle
    \otimes\Big(-a|+\rangle-b|-\rangle\Big),\\
    |\psi''_b\rangle&=&\frac{1}{\sqrt{2}}
    \Big(|+\rangle+|-\rangle\Big)\otimes|+\rangle
    \otimes\Big(-a|+\rangle+b|-\rangle\Big),\\
    |\psi''_c\rangle&=&\frac{1}{\sqrt{2}}
    \Big(|+\rangle-|-\rangle\Big)\otimes|-\rangle
    \otimes\Big(a|+\rangle+b|-\rangle\Big),\\
    |\psi''_d\rangle&=&\frac{1}{\sqrt{2}}
    \Big(|+\rangle+|-\rangle\Big)\otimes|-\rangle
    \otimes\Big(a|+\rangle-b|-\rangle\Big).
\end{eqnarray}
\end{subequations}

\item [Step 5, 6 and 7 -- Local Detection, Classical communication and Rotation:]

(These last three steps are only briefly described here. For more
details, see for instance Lomonaco \cite{Lomonaco1}). {\bf Step
5:} Once the system is in a product state, Alice is supposed to be
able to determine by means of local detectors which state is
actually the state of the system among the states
$|\psi''_\alpha\rangle$. {\bf Step 6:} Alice communicates her
results to Bob by classical means. {\bf Step 7:} Hence, knowing in
which of the above states is particle 3 accordingly, Bob can
perform a rotation in order to have his particle 3 in the state
initially carried by particle 1.
\end{description}
\subsection{Refutation of this Quantum Teleportation
Scheme}\label{ref}
The point of this section is to demonstrate in
the light of Sects. \ref{math} and \ref{Appl} that although
absolutely essential to the Quantum Teleportation process, {\bf
Step 3} cannot be performed by any measurement apparatus as it is
actually irrelevant to Quantum Mechanics.

First, it is important to stress the distinction between {\bf Step
2}, an algebraic operation which doesn't change the physics behind
the vector state---as Eq. (\ref{init}) is then simply rewritten in
a more suggestive way---and {\bf Step 3}, a physical measurement
which strongly change the nature of the state---from the pure
state of Eq. (\ref{TelEq1}) to the mixture of states of Eq.
(\ref{density1}).

The point is that {\bf Step 3} is absolutely essential to complete
a Quantum Teleportation scheme, that is, {\bf Step 2} and {\bf
Step 4} alone are not enough to guarantee the completion of the
quantum teleportation process. Indeed, if the conditional spin
flip $\hat{U}_\mathrm{C}$ is applied ({\bf Step 4}) directly on
the initial \emph{pure} state of Eq.(\ref{TelEq1}) written in the
Bell states basis ({\bf Step 2}) instead of the \emph{mixture} of
Eq. (\ref{density1}), as is sometimes suggested in the literature
\cite{Lomonaco1}, then the result is radically different than the
desired teleportation:
\begin{eqnarray}\label{TelEq4}
    |\widetilde{\psi}'_{123}\rangle&=&(\hat{U}_C\otimes\hat{1\negmedspace\mathrm{l}}_3)
  (\hat{1\negmedspace\mathrm{l}}_\mathrm{Bell}\otimes\hat{1\negmedspace\mathrm{l}}_3)
  |\psi_{123}\rangle \\
  &=&(\hat{U}_C\;\hat{1\negmedspace\mathrm{l}}_\mathrm{Bell}
  \otimes\hat{1\negmedspace\mathrm{l}}_3)|\psi_{123}\rangle \\
  &=&(\hat{U}_C\otimes\hat{1\negmedspace\mathrm{l}}_3)|\psi_{123}\rangle,
\end{eqnarray}
as it is easy to check that
$\hat{U}_C\;\hat{1\negmedspace\mathrm{l}}_\mathrm{Bell}=\hat{U}_C$,
and therefore:
\begin{equation}\label{wrongtel}
    |\widetilde{\psi}'_{123}\rangle=
    a|+\rangle\otimes|\Phi^-_{23}\rangle
    +b|-\rangle\otimes|\Psi^-_{23}\rangle.
\end{equation}
This is but a conditional spin flip applied on the initial state
written in the spin states basis, not at all a teleportation
scheme. Note that this problem does not arise when $\hat{U}_C$ is
applied on the mixture $\rho'_{123}$ (i.e., after {\bf Step 3}),
as is done in Eq. (\ref{transdens}), since the conditional spin
flip $(\hat{U}_C\otimes\hat{1\negmedspace\mathrm{l}}_3)$ is then
by no means applied on $|\psi_{123}\rangle$ but on the
$|\psi'_\alpha\rangle$ states instead, leading indeed to a
teleportation scheme.

Therefore, being able to discriminate between the four Bell states
-- or even to discriminate just one Bell state
\cite{Braunstein2,Bouwmeester1} -- requires that the system hold
by Alice is \emph{already} in one of the four Bell states, which
is not enough to perform a quantum teleportation: Alice must first
find a way to reduce the pure state of Eq. (\ref{TelEq1}) into the
mixture of Eq. (\ref{density1}).

To perform the reduction of {\bf Step 3}, a measurement
represented by \emph{an observable having the four Bell states for
eigenvectors} must be performed on the system described by the
pure state $|\psi_{123}\rangle$. Indeed, according to the von
Neumann reduction postulate, if a measurement represented in
$\mathcal{H}_1\otimes\mathcal{H}_2$ by an observable having the
four Bell states for eigenvectors is performed on the system
$|\psi_{123}\rangle$, then one particular projection
$\hat{P}_\alpha$ on the Bell states
\begin{subequations}
    \label{TelEq2a}
\begin{eqnarray}
    \hat{P}_a&=&|\Psi^-\rangle\langle\Psi^-|\otimes\hat{1\negmedspace\mathrm{l}}_3,\\
    \hat{P}_b&=&|\Psi^+\rangle\langle\Psi^+|\otimes\hat{1\negmedspace\mathrm{l}}_3,\\
    \hat{P}_c&=&|\Phi^-\rangle\langle\Phi^-|\otimes\hat{1\negmedspace\mathrm{l}}_3,\\
    \hat{P}_d&=&|\Phi^+\rangle\langle\Phi^+|\otimes\hat{1\negmedspace\mathrm{l}}_3,
\end{eqnarray}
\end{subequations}
occurs with the probability
\begin{equation}\label{proba1}
  \mathcal{P}_\alpha=\langle\psi_{123}|\hat{P}_\alpha|\psi_{123}\rangle=\frac{1}{4},
\end{equation}
and the corresponding final state of the whole system is
\begin{equation}\label{project1}
  |\psi'_\alpha\rangle=
  \frac{\hat{P}_\alpha|\psi_{123}\rangle}{\sqrt{\langle\psi_{123}|\hat{P}_\alpha|\psi_{123}\rangle}},
\end{equation}
which is one of the state of Eqs. (\ref{TelEq2}), so that the
system is indeed represented by the mixture of Eq.
(\ref{density1}).

In the literature, the well known operator supposed to achieve
this task is the Bell operator \cite{Braunstein1}, defined as
\begin{equation}\label{Bellop}
  \hat{B}_\mathrm{CHSH}\equiv
  \sqrt{2}\big(\sigma_x\otimes\sigma_x+\sigma_z\otimes\sigma_z\big).
\end{equation}
Of course, as long as $\sigma_x\otimes\sigma_x$ and
$\sigma_z\otimes\sigma_z$ were understood as commuting observables
sharing the Bell states as eigenvectors (as was the case in Sect.
\ref{Bellcode}, and always the case in the literature), the Bell
operator, which is a linear combination of these observables,
appeared as an observable as well with the Bell states for
eigenvectors. However, as seen in Sect. \ref{Bellcode}, these
predictions are based on the use of correlation degeneracy and are
therefore incorrect. It was demonstrated on the contrary in Sect.
\ref{Belljoei} that:
\begin{itemize}
  \item \emph{the Bell states are not eigenvectors of
$\sigma_x\otimes\sigma_x$ and $\sigma_z\otimes\sigma_z$}.
  \item \emph{$\sigma_x\otimes\sigma_x$ and $\sigma_z\otimes\sigma_z$ are
not commuting observables}.
\end{itemize}
Hence, the Bell states are \emph{a priori} not eigenvectors of the
Bell operator, and as a linear combination of non commuting
observables, the Bell operator is \emph{a priori} not even an
observable. In particular it is definitely not equivalent to a
$\sigma_x\otimes\sigma_x$ measurement followed by a
$\sigma_z\otimes\sigma_z$ measurement, unlike what is sometimes
met in the literature \cite{Cabello1,Larsen1}.

Note that the proposition made by Fujii \cite{Fujii1} to
circumvent the problem raised here with another linear combination
such as $(\sigma_x\otimes\sigma_x+2\sigma_z\otimes\sigma_z)$
instead of the Bell operator is of no help: the problem arises for
any linear combinations of $\sigma_x\otimes\sigma_x$ and
$\sigma_z\otimes\sigma_z$, since these observables do not commute
and since their eigenvectors are not the Bell states.

More generally, discarding the use of correlation degeneracy --
which is made necessary by the wrong predictions of Sects.
\ref{entangeig} and \ref{Bellcode}, implies that the mere
possibility of an observable having the Bell states for
eigenvectors is physically meaningless, as it requires the very
use of correlation degeneracy and all its corollary of wrong
predictions. In other words, \emph{\mbox{no observable has the
Bell states for eigenvectors}}. Therefore, {\bf Step 3} is
irrelevant to Quantum Mechanics, and no measurement is capable of
triggering the reduction of the pure state of Eq. (\ref{TelEq1})
into the mixture of Eq. (\ref{density1}). Without this essential
{\bf Step 3}, the ideal scheme described in Sect. \ref{QTscheme}
is therefore incomplete, and cannot to lead to any Quantum
Teleportation.
\section{Conclusion}
The inconsistency pointed out within Quantum Mechanics between the
predictions derived from the use of correlation degeneracy (in
Sect. \ref{entangeig}) and the predictions derived from the
quantum postulates (in Sect. \ref{phys}) was solved by the
introduction of joint eigenvalue representation (in Sect.
\ref{jointeigen}). The use of entangled states as eigenvectors of
tensor product observables was thus discarded. Applied to the Bell
states (in Sect. \ref{Appl}), these results led to rather
different predictions than those usually encountered in the
literature. It was shown as a consequence (in Sect.
\ref{QTeleport}) that since it requires the use of such entangled
vectors as eigenvectors, the concept of Quantum Teleportation is
therefore irrelevant to Quantum Mechanics.
\acknowledgments I am
indebted to Werner Hofer, Kim Kirkpatrick, Al Kracklauer, Norbert
Lutkenhaus, Daniel Oi, Tony Short, Roger Schafir, and Shawn Pethel
for very helpful comments on the first version of this article.

\end{document}